\documentclass[11pt]{article}
\usepackage{graphicx}
\usepackage{latexsym,graphicx}
\usepackage{float}
\usepackage{amsmath}
\usepackage{amsfonts}
\usepackage{amssymb}
\usepackage{epstopdf}
\DeclareGraphicsExtensions{.eps}
\usepackage[left=2.5cm,top=2.5cm,right=2.5cm,bottom=2.5cm]{geometry}

\catcode `\@=11
\catcode `\@=12
\begin{document}
	\begin{center}
		\large{\bf{Quintessence Behaviour of an Anisotropic Bulk Viscous Cosmological Model in Modified $f(Q)$-Gravity}} \\
		\vspace{5mm}
		\normalsize{Anirudh Pradhan$^1$, Archana Dixit$^2$, Dinesh Chandra Maurya$^3$}\\
		\vspace{5mm}
		\normalsize{$^{1}$ Centre for Cosmology, Astrophysics and Space Science, GLA University, Mathura-281 406,
		Uttar Pradesh, India}\\
		\vspace{2mm}
		\normalsize{$^{2}$Department of Mathematics, Institute of Applied Sciences and Humanities, GLA University, Mathura-281 406,
		Uttar Pradesh, India}\\
		\vspace{2mm}
		\normalsize{$^3$Center for Theoretical Physics and Mathematics, IASE (Deemed to be University), Sardarshahar-331 403 (Churu),
		Rajsthan, India. }\\
		\vspace{2mm}
		$^1$E-mail:pradhan.anirudh@gmail.com\\	
		\vspace{2mm}
		$^2$E-mail:archana.dixit@gla.ac.in\\
		\vspace{2mm}
		$^3${Email:dcmaurya563@gmail.com}\\
		
	\end{center}
	\vspace{5mm}
\begin{abstract}
In this article, we have discussed the results of our investigation. We consider an anisotropic viscous cosmological model of (LRS) Bianchi type I spacetime universe filled with a viscous fluid under $f(Q)$ gravity. We have studied the modified $f(Q)$ gravity with quadratic form $f(Q)=\alpha Q^{2}+\beta$, where $Q$ is called as a non-metricity scalar and $\alpha $, $\beta$ are positive constants. We obtain the modified Einstein field equation by considering the  viscosity coefficient $\xi(t)=\xi_{0}H$ and obtained the scale factor $a(t)=2~sinh\left (\frac{m+ 2} {6}\sqrt{\frac{\xi_{0}}{\alpha(2m+1)}}t\right)$.We applied the observational constraint on the apparent magnitude $m(z)$ using the $\chi^{2}$ test formula with observational  data set like JLA or Union 2.1 compilation and obtained the best approximate values of the model parameters ${ m,\alpha, H_{0} \xi_{0}}$.
We have analyzed our model and found a transit phase quintessence anisotropic accelerating universe. We also examined the bulk viscosity equation of state (EoS) parameter $\omega_{v}$ and obtained its current value as $\omega_{v}<-1/3$, which shows the dark energy dominant model, cosmological constant, phantom, and super-phantom dark energy models, and tends to the $\Lambda$CDM value ($\omega_{v}=-1$) in the late time. We also estimate the current age of the universe as $t_{0}\approx13.6$ Grys and analyze the Statefinder parameters with $(s,r)\to(0,1)$ as $t \to \infty$.

\end{abstract}
\vspace{5mm}


\section{Introduction}
Cosmological measurements show that the universe's expansion is accelerating \cite{ref1}. There has been a lot of recent research on its potential mechanisms, including extended gravity \cite{ref2}, the decaying Cold Dark Matter (CDM) model \cite{ref3}, the modification of the equation of state (hereafter EoS), or the introduction of various types of the so-called dark energy models to account for the observed cosmic acceleration expansion.
It is well known from the ``recent astrophysical observations that the present-day universe is undergoing an accelerated expansion phase due to an unknown mystical thing that has a large negative pressure termed Dark Energy (DE)" \cite{ref4, ref5}. Further, the observations of WMAP satellite \cite{ref6}, and large-scale structure \cite{ref7} measured the cosmic microwave background (CMB) anisotropies and suggested that approximately $70\%$ of the universe be made up of dark energy (DE), and the rest is made up of relativistic dark matter (DM), and ordinary matter \cite{ref8}.
The nature of  (DE) has previously been classified using the equation of state (EoS) parameter $\omega$, which is defined as the ratio of spatially homogeneous pressure $(p)$ to dark energy density $(\rho)$, or $\omega=\frac{p}{\rho}$. Recent cosmological findings show that the $\omega < -1/3$ value of the EoS parameter is a need for the acceleration of cosmic expansion. The essential possibilities in this categorization are scalar field models with an EoS value of $-1<\omega<-1/3$, which articulates as a Quintessence field dark energy \cite{ref9, ref10}, and $\omega<-1$ as a phantom field dark energy \cite{ref11}. For example, the phantom field of DE, has gotten much interest because of its bizarre qualities.
The Phantom Model shows how dark energy grows in an exciting future spread that leads to a future singularity in a finite amount of time. It also violates the four energy constraints that serve to keep wormholes in check \cite{ref12}. In addition, the combined findings of the $H_{0}$ measurements, WMAP9 \cite{ref13},  ``SNe-Ia, cosmic microwave background, and BAO demonstrate that $\omega=-1.084\pm0.063$ for dark energy. This is despite the fact that the Planck team demonstrated in 2015 that $\omega=11.006\pm0.0451$ \cite{ref14}". Cao et al. \cite{ref14a} demonstrated analytically the integrability of the dynamics of the Kerr-Newman black hole is surrounded by electromagnetic field, quintessence, and charged particles moving around it for a closed FRW cosmology model with a conformally coupled scalar field. Wu  \cite{ref14b} studied a new interpretation of zero Lyapunov exponents in Belinskii-Khalatnikov-Lifshitz time for Mixmaster cosmology. Ma et al. \cite{ref14c}  computationally investigated the impact of changing the cosmological constant $\Lambda$ and self-interacting coefficient $\lambda$ on the transition to chaos in complexified space.
\\

Inspired by studies on black hole thermodynamics \cite{ref15}, Hooft presented the renowned holographic principle (HP) for the first time \cite{ref16}. Later, the HP was applied to the DE problem, resulting in a new model of DE known as the Holographic Dark Energy (HDE), in which the energy density is based on physical quantities on the universe's boundary, such as the ``reduced Planck mass and the cosmological length scale", which is chosen as the universe's future event horizon  \cite{ref17}. Many authors have studied the development of different cosmological models that include viscosities in their fluids. The viscosity theory of relativistic fluids was first proposed by Eckart  \cite{ref18} and Landau and Lifshitz \cite{ref19}, who considered only the first order of change from equilibrium. Israel \cite{ref20} came up with the idea for the relativistic second-order theory, and Israel and Stewart \cite{ref21} worked on it together. Similarly Murphy \cite{ref22} found that a zero-curvature cosmological model is perfectly solvable even in the presence of bulk viscosity. Huang \cite{ref23} worked on the Bianchi type I cosmological model with bulk viscosity. Many significant problems with the standard Big-Bang cosmology may be resolved by resorting to inflationary cosmology, which is related with the bulk viscosity of the grand unified theory (GUT). \\

Several sources of inspiration for hypotheses go beyond the conventional explanation of gravity. According to Jimenez et al. \cite{ref24}, this result established a new gravity known as symmetric teleparallel gravity or $f(Q)$ gravity. The nonmetricity term, $Q$, is what causes the gravitational interaction. The theory was quickly developed by Lazkoz et al. \cite{ref25} , along with observational constraints to oppose it against the f(Q) Lagrangian as a polynomial function of the redshift, $z$, and an intriguing set of constraints on $f(Q)$ gravity. Recently, Mandal et al. \cite{ref26} published a thorough examination of the energy requirements that allowed for the fixation of free parameters, so limiting the
families of $f(Q)$ models that are consistent with the accelerated expanding behavior of the universe. Also, the $f(Q)$ gravity is the simplest modification of STEGR. Many issues have been discussed in the framework of $f(Q)$ gravity enough to motivate us to work under this new framework. The energy constraints and cosmography in $f(Q)$ theory are examined \cite{ref27,ref28}, using the assumption of a power-law function. Harko et al. \cite{ref29} researched the coupling matter in modified $Q$ gravity. For a $f(Q)$ polynomial model, Dimakis et al. \cite{ref30} addressed quantum cosmology see also \cite{ref31}.\\

The first cosmological solutions in $f(Q)$ gravity can be found in  \cite{ref32, ref33}, but $f(Q)$ cosmography with energy conditions are discussed in \cite{ref19,ref20}. However, as several studies have shown, the profligacy of viscous fluids, including both ``shear and bulk viscosity," may have played an important role in the evolution of the cosmos  \cite{ ref29,ref33, ref38,ref39,ref40}.In reference \cite{ref41}-\cite{ref45e}, the impact of bulk viscosity fluid in the late-time accelerating cosmos was examined. However, in an expanding universe, a feasible process for forming viscous fluids is more difficult to find. The bulk viscosity emerges due to the interruption of the local thermodynamic equilibrium in the context of theoretical study \cite{ref46}.\\

Recent research has focused on the transit phase universe with $f(Q, T)$ gravity \cite{ref47,ref47a},$f(Q)$ gravity \cite{ref48}
Capozziello and Agostino \cite{ref49} examined $f(Q)$-gravity using a model-independent approach and addressed a variety of cosmological characteristics.  Anisotropic $f(Q)$ gravity with bulk viscosity has been examined by Koussour et al. \cite{ref50}, Maurya \cite{ref51}, and Dixit et al. \cite{ref52}. In our current analysis, we will study different mechanisms of DE in the $f(Q)$ gravity model, where $Q$ is the non-metricity scalar responsible for gravity. In Weyl's geometry, the covariant derivative of the metric tensor is non-null. This characteristic can be displayed mathematically in terms of a novel geometric quantity named non-metricity \cite{ref55}. Geometrically, the non-metricity can be described as the variation of the length of a vector during parallel transport. To better understand our Universe, we are obligated to replace the curvature concept with a more general geometrical concept. We consider $f(Q)$ gravity in an anisotropic background and solve the field equations for the average scale factor $a(t)$ and the metric coefficients, which is distinct from what other people have done because most people have just assumed it in the past. Using this scale factor, we've discussed the deceleration parameter as $q, q_{x}, q_{y}$ and the EoS parameter, among other things. \\

The paper is organized as follows:Introduction and an overview of the literature are found in Section 1. We present the $f(Q)$ gravity formalism in Sec. 2 along with the corresponding field equation for LRS Bianchi type-I space-time. In Section 3, we used the bulk-viscosity factor $xi(t)=xi 0H$ and other cosmological solutions to solve Einstein's modified field equations. Sec. 4, devoted to observational constraint. In section 5, we discussed the results, how old the universe is now, and how the statefinder works. In section 6, we have the concluding remarks.


	\section{Evolution of Field Equations in $f(Q)$-gravity}
To find the modified Einstein's field equations in $f(Q)$-gravity, we adopted the action principle mentioned in \cite{ref32,ref33} as
\begin{equation}\label{eq1}
  S=\int\left[\frac{1}{2\kappa}f(Q)+L_{m}\right]dx^{4}\sqrt{-g}
\end{equation}
where $f(Q)$ is an arbitrary function of $Q$ known as non-metricity and other notations are in their usual meaning like matter Lagrangian density $L_{m}$, metric tensor $g_{ij}$ with determinant $g$. The scalar non-metricity $Q$ is expressed as below

\begin{equation}\label{eq2}
Q\equiv -(L^{\alpha}_{\beta i}L^{\beta}_{j\alpha}-L^{\alpha}_{\beta\alpha}L^{\beta}_{ij})g^{ij}~~\text{and}~~Q_{\alpha}={Q_{\alpha}^{i}}_{i}
\end{equation}
where $Q_{\alpha}$ is known as non-metricity tensor and $L^{\alpha}_{\beta\gamma}$ called as deformation tensor and expressed as below

\begin{equation}\label{eq3}
L^{\alpha}_{\beta\gamma}=-\frac{1}{2}g^{\alpha\lambda}(\nabla_{\gamma}g_{\beta\lambda}+\nabla_{\beta}g_{\lambda\gamma}-\nabla_{\lambda}g_{\beta\gamma})
\end{equation}
Now, we obtain the modified field equations after the variation of action (as mentioned in Eq.~(\ref{eq1})) with respect to metric tensor $g_{ij}$, as given below:
\begin{equation}\label{eq4}
  \frac{2}{\sqrt{-g}}\nabla_{\lambda}(\sqrt{-g}f_{Q}P^{\lambda}_{\mu\nu})-\frac{1}{2}fg_{\mu\nu}+f_{Q}
  (P_{\nu\rho\sigma}Q_{\mu}^{\rho\sigma}-2P_{\rho\sigma\mu}Q_{\nu}^{\rho\sigma})=\kappa T_{\mu\nu}
\end{equation}
where $P^{\alpha}_{ij}$ called as supper-potential of the model (as mentioned in \cite{ref32}) and $f_{Q}=\frac{df}{dQ}$.\\
The stress-energy-momentum tensor $T_{\mu\nu}$ is expressed as

\begin{equation}\label{eq5}
T_{ij}=-\frac{2}{\sqrt{-g}}\frac{\delta (\sqrt{-g}L_{m})}{\delta g^{ij}}
\end{equation}
Recently constructed dark energy models with bulk viscosity in $f(Q)$-gravity in \cite{ref48} with $f(Q)=-\alpha Q-\beta$ in a flat FLRW universe and
in this paper, we will derive a viscous dark energy model with arbitrary function $f(Q)=\alpha Q^{2}+\beta$ in LRS Bianchi Type-I space-time metric:
\begin{equation}\label{eq6}
  ds^{2}=A(t)^{2}dx^{2}+B(t)^{2}(dy^{2}+dz^{2})-dt^{2}
\end{equation}
Here $A(t)$ and $B(t)$ are known as metric coefficients and are functions of cosmic time $t$. For the line element (\ref{eq6}), the non-metricity scalar $Q$ is calculated as
\begin{equation}\label{eq7}
  Q=-4\frac{\dot{A}}{A}\frac{\dot{B}}{B}-2\left(\frac{\dot{B}}{B}\right)^{2}
\end{equation}
The stress-energy-momentum tensor for viscous fluid is considered as
\begin{equation}\label{eq8}
  T^{\mu}_{\nu}=diag[-\rho, \tilde{p_{x}}, \tilde{p_{y}}, \tilde{p_{z}}]
\end{equation}
with energy density $\rho$, anisotropic pressures for bulk-viscosity fluid $\tilde{p_{x}}$, $\tilde{p_{y}}$, and $\tilde{p_{z}}$ respectively along $x$, $y$ and $z$ axes. The equation of state parameter for the bulk-viscosity fluid is assuming as $\tilde{p_{i}}=\rho\tilde{\omega_{i}}$ for $i=x, y, z$. Then the diagonal of the stress-energy-momentum tensor becomes
\begin{equation}\label{eq9}
  T^{\mu}_{\nu}=diag[-1, \tilde{\omega_{x}}, \tilde{\omega_{y}}, \tilde{\omega_{z}}]\rho=[-1, \omega_{v}, \omega_{v}+\delta_{v}, \omega_{v}+\delta_{v}]\rho
\end{equation}
where considering EoS parameter along $x$-axis as $\tilde{\omega_{x}}=\omega_{v}$ and taking a deviation $\delta_{v}$ in EoS parameter along $y$ and $z$ axes from the EoS of $x$-axis and is called as skewness parameter. The suffix $(_{v})$ corresponds bulk-viscosity fluid. The EoS parameters $\omega_{v}$ and $\delta_{v}$ may be constant or functions of cosmic time $t$.\\

In a co-moving coordinate system, we can obtain the following field equations from Eqs.~(\ref{eq4}), (\ref{eq6}) and (\ref{eq8}) as given below
\begin{equation}\label{eq10}
  \frac{f}{2}+f_{Q}\left[4\frac{\dot{A}}{A}.\frac{\dot{B}}{B}+2\left(\frac{\dot{B}}{B}\right)^{2}\right]=\rho
\end{equation}
\begin{equation}\label{eq11}
  \frac{f}{2}-f_{Q}\left[-2\frac{\dot{A}}{A}.\frac{\dot{B}}{B}-2\frac{\ddot{B}}{B}-2\left(\frac{\dot{B}}{B}\right)^{2}\right]+
  2\frac{\dot{B}}{B}\dot{Q}f_{QQ}=-\tilde{p_{x}}
\end{equation}
\begin{equation}\label{eq12}
  \frac{f}{2}-f_{Q}\left[-3\frac{\dot{A}}{A}.\frac{\dot{B}}{B}-\frac{\ddot{A}}{A}-\frac{\ddot{B}}{B}-
  \left(\frac{\dot{B}}{B}\right)^{2}\right]+\left(\frac{\dot{A}}{A}+\frac{\dot{B}}{B}\right)\dot{Q}f_{QQ}=-\tilde{p_{y}}=-\tilde{p_{z}}
\end{equation}
Here over dot $(^{.})$ denotes the derivative with respect to time $t$. The volume scale-factor of the considered spacetime universe is defined as
\begin{equation}\label{eq13}
  V=a(t)^{3}=AB^{2}
\end{equation}
with $a(t)$ as the average scale factor and $A,~B$ are the metric coefficients. Now, we define $q$ as deceleration parameter in terms of average scale-factor given by
\begin{equation}\label{eq14}
  q=-\frac{a\ddot{a}}{\dot{a}^{2}}
\end{equation}
The deceleration parameter $(q)$ reveals the various phase of the expanding evolution of the universe, a positive value of $q$ depicts
the decelerating phase and the negative values of $q$ depict the accelerating phase of the universe.\\
Now, the mean Hubble parameter $H$ is defined as
\begin{equation}\label{eq15}
  H=\frac{1}{3}(H_{x}+H_{y}+H_{z})
\end{equation}
with $H_{x}$, $H_{y}$ and $H_{z}$ considered as the directional Hubble parameters respectively along the $x$, $y$ and $z$ axes. For the line element (\ref{eq6}), the directional Hubble parameters are taken as $H_{x}=\frac{\dot{A}}{A}$ and $H_{y}=H_{z}=\frac{\dot{B}}{B}$.\\
Also, we can obtain a connection among the above three parameters $H$, $V$ and $a$ as given below:
\begin{equation}\label{eq16}
  H=\frac{1}{3}\frac{\dot{V}}{V}=\frac{1}{3}\left[\frac{\dot{A}}{A}+2\frac{\dot{B}}{B}\right]=\frac{\dot{a}}{a}
\end{equation}
Now, we define some more geometrical parameters $\theta,~\sigma^{2},~\Delta$ called as scalar expansion, shear scalar and the
mean anisotropy parameter respectively and are given by
\begin{equation}\label{eq17}
  \theta(t)=\frac{\dot{A}}{A}+2.\frac{\dot{B}}{B}
\end{equation}
\begin{equation}\label{eq18}
  \sigma^{2}(t)=\frac{1}{3}\left(\frac{\dot{A}}{A}-\frac{\dot{B}}{B}\right)^{2}
\end{equation}
\begin{equation}\label{eq19}
  \Delta=\frac{1}{3}\sum_{i=x}^{z}\left(\frac{H_{i}-H}{H}\right)^{2}
\end{equation}
with directional Hubble parameters $H_{i}, i=x, y, z$.
\section{Cosmological Solutions}
Now, we have three independent field equations (\ref{eq10})-(\ref{eq12}) in five unknowns $A, B, \rho, \omega_{v}, \delta_{v}$. Therefore, to obtain an exact solution, we are required two more constraints on these parameters, and hence, we assume $(\sigma^{2}\propto\theta^{2})$ the shear is proportional to the expansion scalar, and this relation leads to (see in \cite{ref56})
\begin{equation}\label{eq20}
  A=B^{m}
\end{equation}
with $m\neq1$ as an arbitrary constant and with $m=1$ the model represents an isotropic universe, otherwise, anisotropic universe is obtained.\\
Now, in the present investigation, we have considered quadratic form of the function $f(Q)$ as
\begin{equation}\label{eq21}
  f(Q)=\alpha Q^{2}+\beta, ~~~~~\alpha>0,~~~~\beta>0
\end{equation}
where $\alpha$ and $\beta$ are arbitrary positive constants.\\

Now, using equation (\ref{eq20}) in (\ref{eq13}), the metric coefficients are obtained in terms of scale factor $a(t)$ as:
\begin{equation}\label{eq22}
  A=a(t)^{\frac{3m}{m+2}}, \hspace{1cm} B=a(t)^{\frac{3}{m+2}}
\end{equation}
and Hubble components are given by
\begin{equation}\label{eq23}
  H_{x}=\frac{\dot{A}}{A},~~~~H_{y}=H_{z}=\frac{\dot{B}}{B}=\frac{1}{m}\frac{\dot{A}}{A}=\frac{1}{m}H_{x}
\end{equation}
Viscous fluid pressures are defined as \cite{ref19,ref50} in the $x$, $y$, and $z$ directions.
\begin{equation}\label{eq24}
\tilde{p_{x}}=p-3H_{x}\xi(t)\hspace{1cm} \tilde{p_{y}}=p-3H_{y}\xi(t)\hspace{1cm}\tilde{p_{z}}=p-3H_{z}\xi(t)
\end{equation}
where $p$ is normal pressure and $\xi$ is formed when the viscous fluid deviates from local thermal equilibrium.
$\xi$ can be a function of the Hubble parameter and its derivative \cite{ref19,ref58}.\\
Subtracting (\ref{eq12}) from (\ref{eq11}) using the expression ~(\ref{eq24}) gives
\begin{equation}\label{eq25}
f_{Q}\left[\frac{\dot{A}}{A}\frac{\dot{B}}{B}+\frac{\ddot{A}}{A}-\frac{\ddot{B}}{B}-\left(\frac{\dot{B}}{B}\right)^{2}\right]+
\left(\frac{\dot{A}}{A}-\frac{\dot{B}}{B}\right)\dot{Q}f_{QQ}+3\xi(t)(H_{x}-H_{y})=0
\end{equation}
From the expression (\ref{eq21}) ,
\begin{equation}\label{eq26}
  f_{Q}=2\alpha Q, \hspace{1cm} f_{QQ}=2\alpha
\end{equation}
Using Eq.~(\ref{eq23}) with (\ref{eq7}) gives the non-metricity scalar as
\begin{equation}\label{eq27}
  Q=-\frac{2(2m+1)}{m^{2}}H_{x}^{2}
\end{equation}
Using Eq.~(\ref{eq24}) for the viscous universe of Eq.~(\ref{eq10})-(\ref{eq12}), the parameter bulk viscosity coefficient $\xi$ is not a problem. , being the Hubble parameter and its derivative, we assume $\xi=\xi(H)$ and consider the special form of $\xi$ as \cite{ref59}-\cite{ref67}.
\begin{equation}\label{eq28}
  \xi(t)=\xi_{0}H
\end{equation}
where $\xi_{0}$ is an arbitrary constant.\\
Now, using Eqs.~(\ref{eq26}) to (\ref{eq28}) in (\ref{eq25}), we get
\begin{equation}\label{eq29}
  \dot{H_{x}}+\frac{(m+2)}{3m}H_{x}^{2}-\frac{\xi_{0}m(m+2)}{12\alpha(2m+1)}=0
\end{equation}
Solving Eq.~(\ref{eq29}), we get the component $H_{x}$ of Hubble parameter $H$ as
\begin{equation}\label{eq30}
  H_{x}=\frac{m}{2}\sqrt{\frac{\xi_{0}}{\alpha(2m+1)}}coth\left(\frac{m+2}{6}\sqrt{\frac{\xi_{0}}{\alpha(2m+1)}}t\right)
\end{equation}
and hence, $H_{y}$ is given by
\begin{equation}\label{eq31}
  H_{y}=\frac{1}{2}\sqrt{\frac{\xi_{0}}{\alpha(2m+1)}}coth\left(\frac{m+2}{6}\sqrt{\frac{\xi_{0}}{\alpha(2m+1)}}t\right)
\end{equation}
Now, Integrating Eqs.~(\ref{eq30}) and (\ref{eq31}), we get the metric coefficients as
\begin{equation}\label{eq32}
  A(t)=\left[2sinh\left(\frac{m+2}{6}\sqrt{\frac{\xi_{0}}{\alpha(2m+1)}}t\right)\right]^{\frac{3m}{m+2}}
\end{equation}
\begin{equation}\label{eq33}
  B(t)=\left[2sinh\left(\frac{m+2}{6}\sqrt{\frac{\xi_{0}}{\alpha(2m+1)}}t\right)\right]^{\frac{3}{m+2}}
\end{equation}
Hence, the average scale factor $a(t)$ and the average Hubble parameter $H(t)$ are calculated respectively as
\begin{equation}\label{eq34}
  a(t)=2sinh\left(\frac{m+2}{6}\sqrt{\frac{\xi_{0}}{\alpha(2m+1)}}t\right)
\end{equation}
and
\begin{equation}\label{eq35}
  H(t)=\frac{m+2}{6}\sqrt{\frac{\xi_{0}}{\alpha(2m+1)}}coth\left(\frac{m+2}{6}\sqrt{\frac{\xi_{0}}{\alpha(2m+1)}}t\right)
\end{equation}

The $q_{x}$ and $q_{y}$ components of the deceleration parameter $q(t)$ are computed respectively as

\begin{equation}\label{eq36}
  q_{x}=-1+\frac{1}{3}\frac{m+2}{m}sech^{2}\left(\frac{m+2}{6}\sqrt{\frac{\xi_{0}}{\alpha(2m+1)}}t\right)
\end{equation}
\begin{equation}\label{eq37}
  q_{y}=-1+\frac{1}{3}(m+2)sech^{2}\left(\frac{m+2}{6}\sqrt{\frac{\xi_{0}}{\alpha(2m+1)}}t\right)
\end{equation}
The average deceleration parameter $q(t)$ is calculated as
\begin{equation}\label{eq38}
  q=-1+sech^{2}\left(\frac{m+2}{6}\sqrt{\frac{\xi_{0}}{\alpha(2m+1)}}t\right)
\end{equation}
We now have the energy density $\rho$ from the field equation (\ref{eq10}) to (\ref{eq12}), the equation of state (EoS) $\omega_{v}$ and the skewness parameter $ \delta_{v} $ for bulk viscosity fluid are obtained as
\begin{equation}\label{eq39}
  \rho(t)=\frac{\beta}{2}-\frac{6\alpha(2m+1)^2}{m^4}H_{x}^{4}
\end{equation}
and
\begin{equation}\label{eq40}
  \omega_{v_{x}}=-1+\frac{\left[\frac{\xi_{0}^{2}(m+2)}{6\alpha m(2m+1)}coth^2\left(\frac{m+2}{6}\sqrt{\frac{\xi_{0}}{\alpha(2m+1)}}t\right)-\frac{\xi_{0}^{2}(4m-1)}{2\alpha (2m+1)}coth^4\left(\frac{m+2}{6}\sqrt{\frac{\xi_{0}}{\alpha(2m+1)}}t\right)\right]}{\frac{\beta}{2}-\frac{6\alpha(2m+1)^2}{m^4}H_{x}^{4}}
\end{equation}
and
\begin{equation}\label{eq41}
  \delta_{v_{x}}=-\frac{\frac{4\alpha(2m+1)(m-1)}{m^{4}}[(m+2)H_{x}^{2}+3m\dot{H_{x}}]H_{x}^{2}}{\frac{\beta}{2}-\frac{6\alpha(2m+1)^2}{m^4}H_{x}^{4}}
\end{equation}
\section{Observational Constraints}
To validate the theoretical model, it is necessary to compare the results of derived the model with observational universe models, and the technique of the curve-fitting of some cosmological parameters with various observational data sets is the best way to compare and validation of the model. Hence, we will find the best fit curve of Hubble parameter and apparent magnitude, and performing these curve-fittings, we will have to require a relation between cosmic age and redshift which is taken as the relation between the redshift and average scale-factor as $\frac{a_{0}}{a}=1+z$ and with the standard convention taking $a_{0}=1$. Now, using this relationship, we derived the cosmic age in terms of redshift $z$ as
\begin{equation}\label{eq42}
  t(z)=\frac{6}{m+2}\sqrt{\frac{\alpha(2m+1)}{\xi_{0}}}sinh^{-1}\left(\frac{1}{2(1+z)}\right)
\end{equation}
And hence, the Hubble parameter $H(z)$ in terms of redshift $z$ can be derived as
\begin{equation}\label{eq43}
  H(z)=\frac{m+2}{6}\sqrt{\frac{\xi_{0}}{\alpha(2m+1)}}\sqrt{1+4(1+z)^{2}}
\end{equation}
Several observational datasets are now available on the Internet. Here we use two datasets of the apparent magnitude $m(z)$ using redshift $z$ in the derived model. The first is the $51$ data set of apparent magnitude $m(z)$ from Joint Light Curve Analysis (JLA) \cite{ref68} and  $518$ data set of magnitude $m(z)$ from Union 2.1 Compilation of SNe Ia datasets \cite{ref69}.
\subsection*{Apparent magnitude}
Luminous distance is measured by the total luminous flux of a light source and defined as $D_{L}=(z+1)c\int_{0}^{z}\frac{dz}{H(z)} $. Define the apparent magnitude with respect to $D_{L}$ to obtain the best values of the model parameters $\{\alpha, m, \xi_{0}, H_{0}\}$ for the best-fit curve of $m(z)=16.08+5 log_{10}(\frac{H_{0}D_{L}}{0.026 c\text{Mpc}})$.
Using the $51$ observed apparent magnitude data set $m(z)$ from ``Joint Light Curve Analysis (JLA)" as in \cite{ref68} and the $518$ observed apparent magnitude data set Union 2.1 compiled use like \cite{ref69} from to determine the best-fit curve for the apparent magnitude $m(z)$ using the SNe Ia dataset. We use the $\chi^{2}$ test formula to achieve the best curve fit for theoretical and empirical results:\\

The Luminosity Distance $D L$ is calculated by the formula

\begin{equation}\label{eq44}
D_{L}=c(1+z)\int_{0}^{z}\frac{dz}{H(z)}
\end{equation}
where $c$ is the velocity of light and $H(z)$ is the Hubble parameter given in Eq. (\ref{eq43}).\\
Now, using Eq. (\ref{eq43}) in (\ref{eq44}), we have obtained the luminosity distance $D_{L}$ for the above derived model as
\begin{equation}\label{eq45}
D_{L}=\frac{3c(1+z)}{m+2}\sqrt{\frac{\alpha(2m+1)}{\xi_{0}}}\left[sinh^{-1}(2z+2)-sinh^{-1}(2)\right]
\end{equation}
Consequently, the apparent magnitude $m(z)$ is calculated as
\begin{equation}\label{eq46}
m=16.08+5\log_{10}\frac{H_{0}D_{L}}{0.026cMpc}
\end{equation}

We employed the $\chi2$  test to identify the $m(z)$ curve that best fit the data as;
\begin{equation}\nonumber
\chi^{2}_{SN}=\sum_{i=1}^{N}\left(\frac{(m_{i})_{ob}-(m_{i})_{th}}{(m_{i})_{th}}\right)^{2}
\end{equation}
where summation is taken from $1$ to $51$ data points for the JLA data sets and it taken from $1$ to $518$ data points for union 2.1
compilation data sets.\\

\begin{table}[H]
  \centering
  \begin{tabular}{|c|c|c|c|c|c|c|}
     \hline
     Data & $m$ & $\alpha$ & $\xi_{0}$ & $H_{0}$ & $\chi^{2}$ & $R^2$ \\
     \hline
     JLA       & 2 & $9.52698\times10^{-5}\pm3.34447\times10^{-6}$ & 0.99998 & 70.00056 & 0.07228 & 0.92346 \\
     Union 2.1 & 2 & $8.42571\times10^{-5}\pm8.3134\times10^{-7}$ & 0.97455 & 70.00000 & 0.05915 & 0.99404 \\
     \hline
   \end{tabular}
  \caption{The model parameters that best fit the JLA and Union 2.1 data sets.}\label{Table 1}
\end{table}
\begin{figure}[H]
	a.\includegraphics[width=9cm,height=8cm,angle=0]{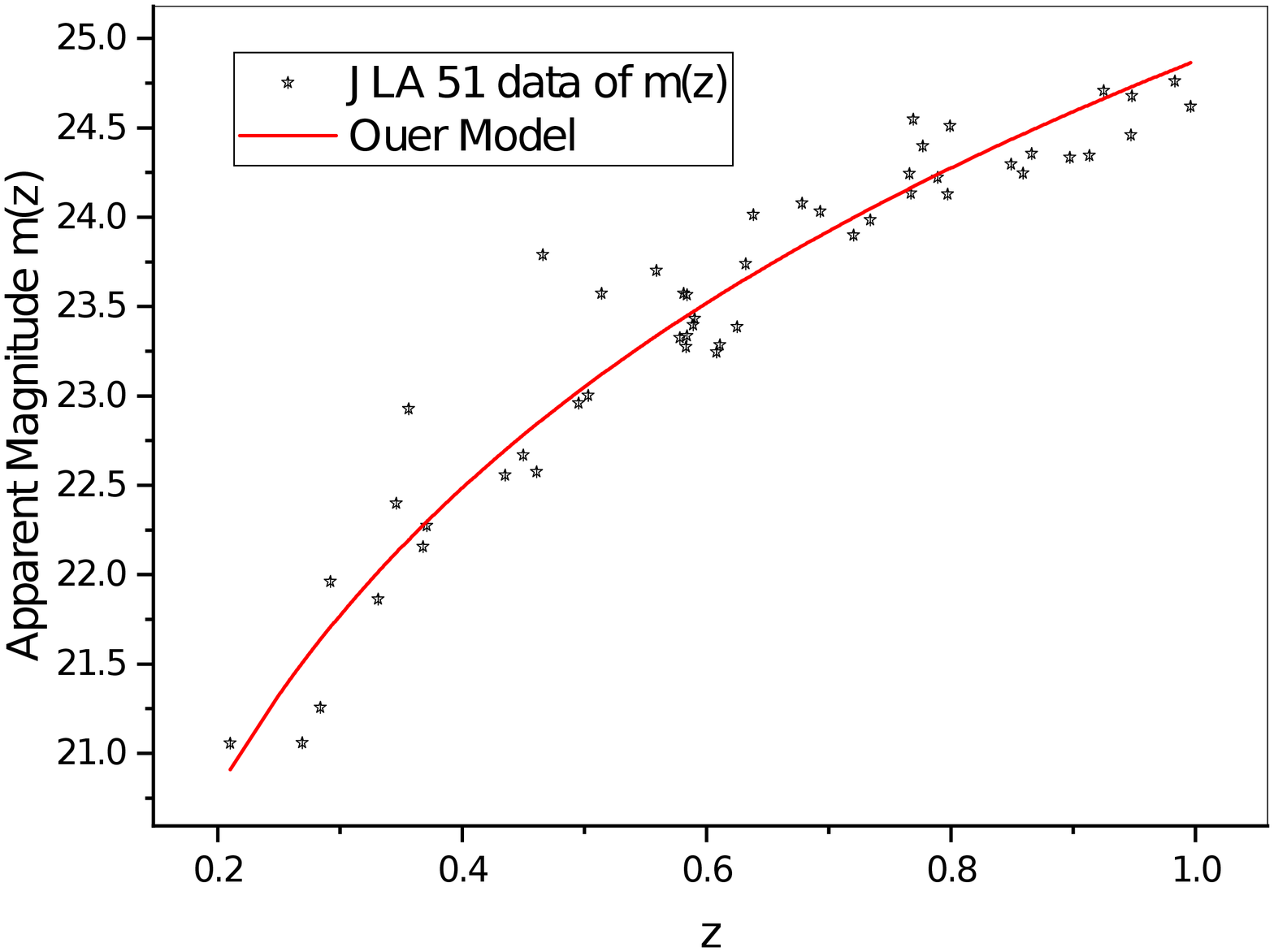}
	b.\includegraphics[width=9cm,height=8cm,angle=0]{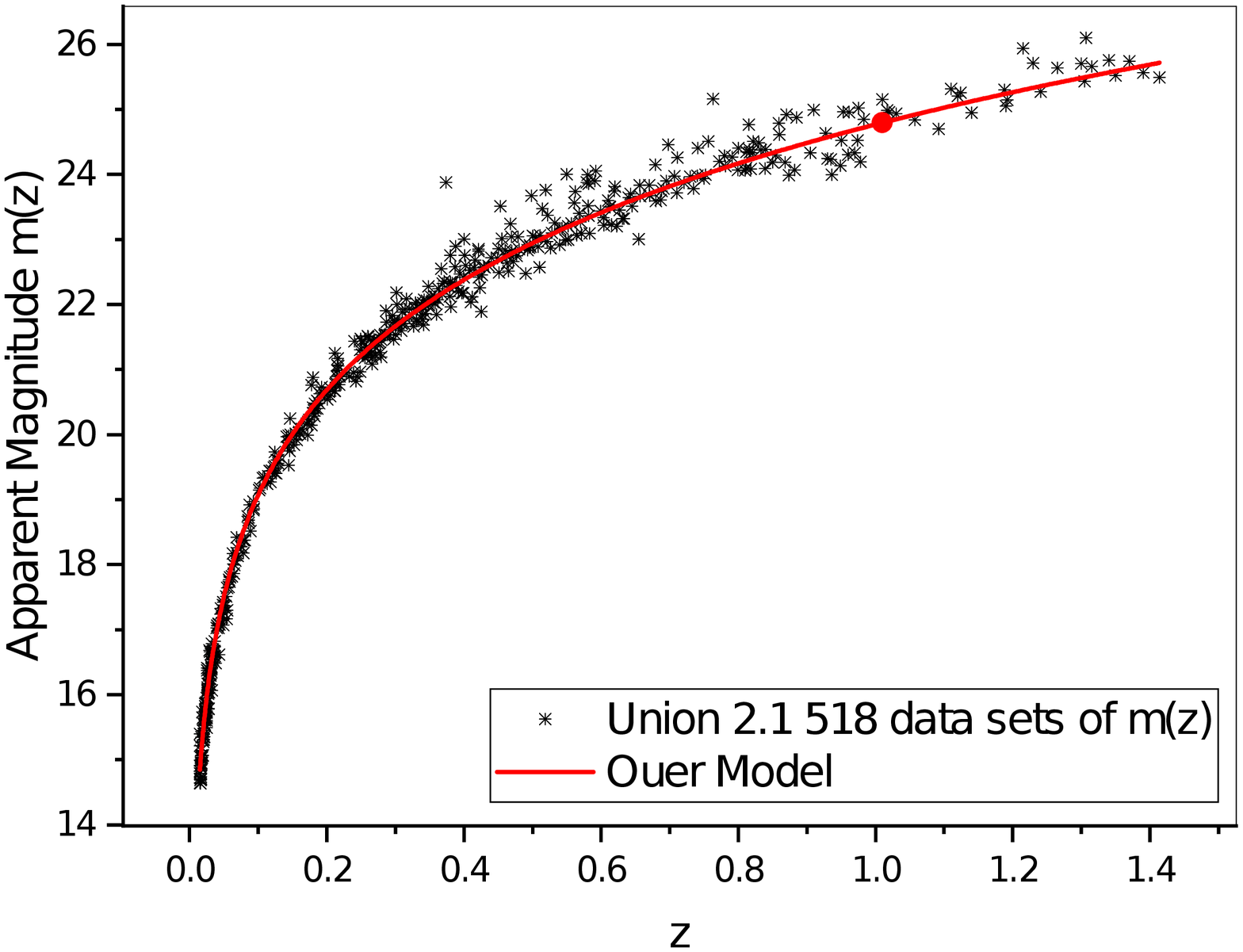}
	\caption{JLA and Union 2.1 data sets, respectively, were used to fit the m(z) curve.
	}
\end{figure}
\begin{figure}[H]
\centering
	\includegraphics[width=14cm,height=10cm,angle=0]{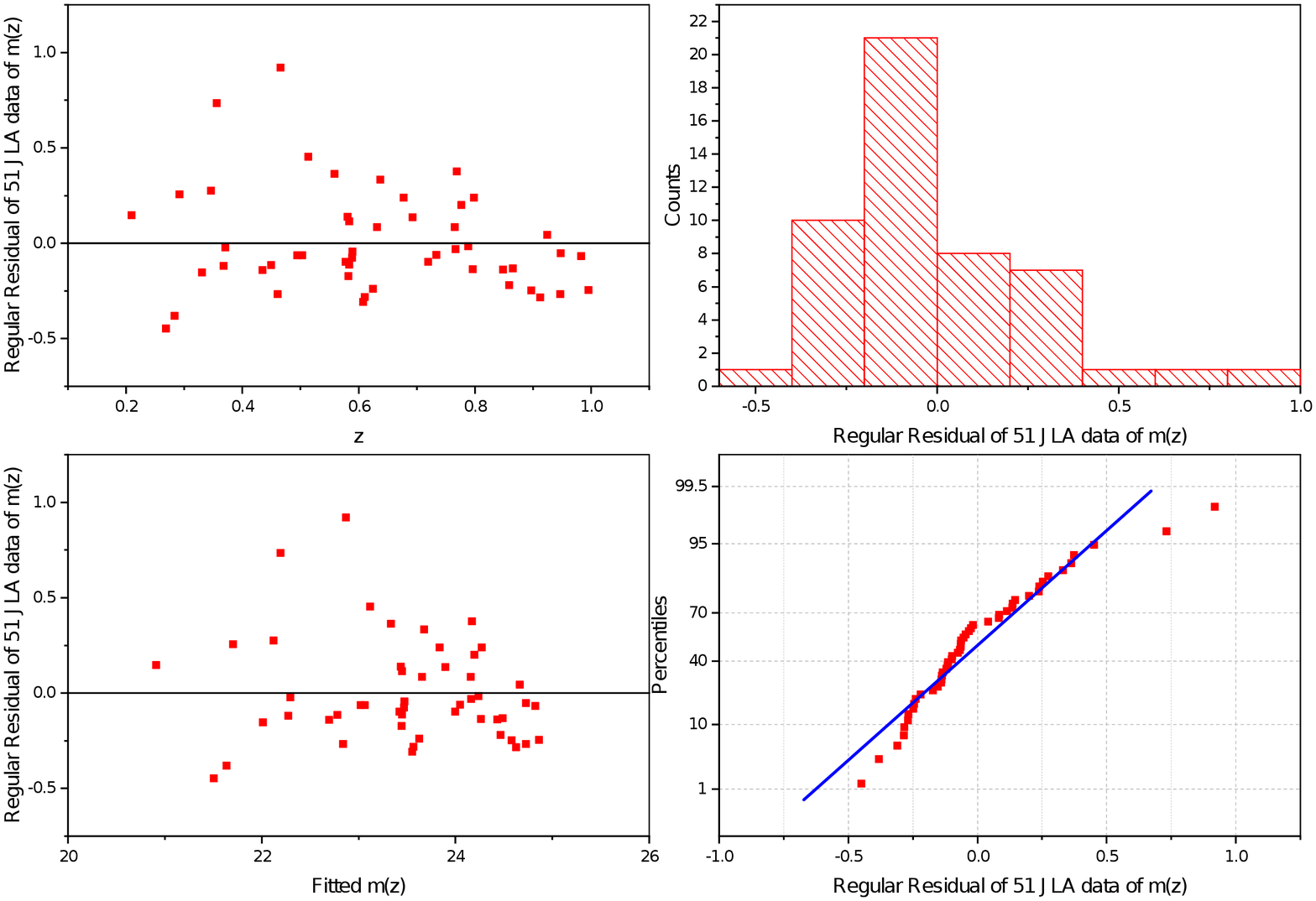}
	\caption{The residual plot of the best fit model and JLA data sets.}
\end{figure}
\begin{figure}[H]
\centering
	\includegraphics[width=14cm,height=10cm,angle=0]{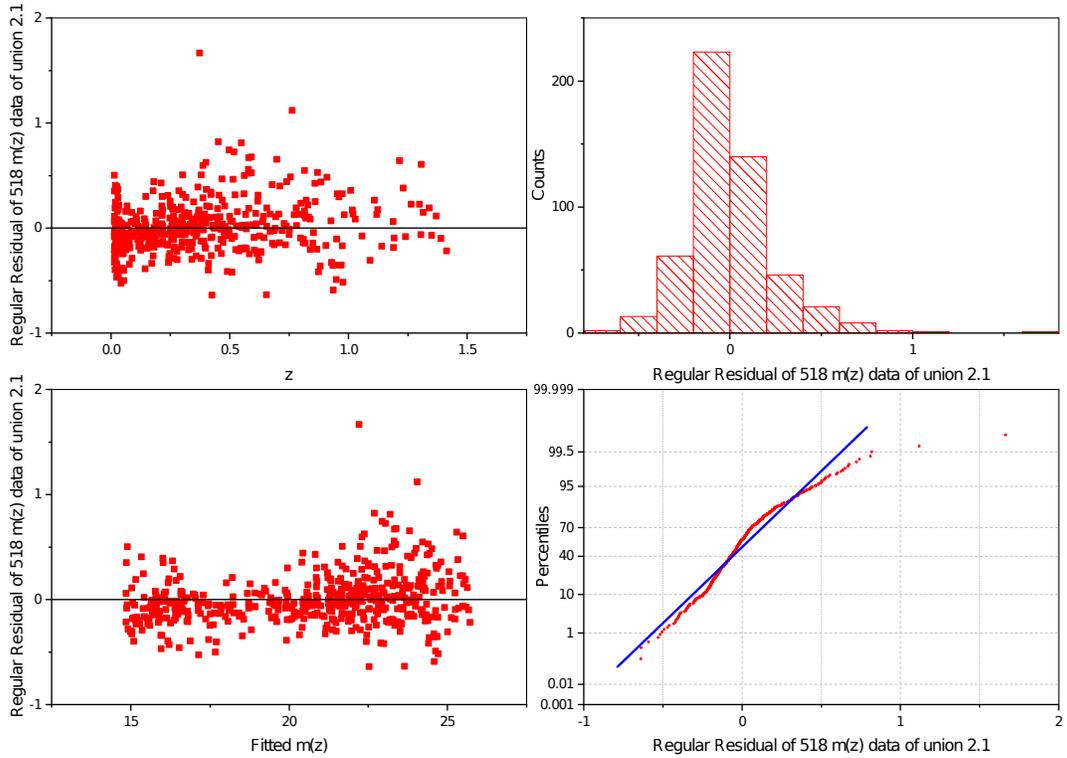}
	\caption{The residual plot of the best fit model and Union 2.1 data sets.}
\end{figure}
With this understanding of $\chi^{2}$  values, we used the $\chi^{2}$-test formula to obtain the best fit curve of apparent magnitude $m(z)$ with observational data sets JLA and union 2.1, as well as the best fit values of the model parameters $\{\alpha, m, \xi_{0}, H_{0}\}$, which are listed in Table 1. The $\chi^{2}=0$ means that observational and theoretical values are exactly the same. Figures 1a and 1b show the best-fit curves of $m(z)$. Figures 2 and 3 demonstrate, using two different sets of data, the accuracy of our fitted model.

\section{Discussion of Results}
The equation (\ref{eq42}) represents a relationship between cosmic time $t$ and redshift $z$, and its geometrical interpretation is given by
 Figure 4a, from which we can say that as $z\to\infty$, the cosmic time $t\to0$. At $z=0$, we find the present value of $t(0)=\frac{6}{m+2}\sqrt{\frac{\alpha(2m+1)}{\xi_{0}}}sinh^{-1}\left(\frac{1}{2}\right)$.
From the Figure 4a, we can see that the cosmic time varies as $0 \leq t \leq 0.014$ over the redshift $0\leq z \leq \infty$. The expression for Hubble parameter $H(z)$ is  represented by the Eq. (\ref{eq42}) and Figure 4b
shows its geometrical behavior. From Figure 4b, we can see that $H$ is an increasing function of redshift $z$, and at $z=0$, the value of $H$
denotes the present value of Hubble parameter $H_{0}$. Hence, at $z=0$, $H_{0}=\frac{m+2}{6}\sqrt{\frac{5\xi_{0}}{\alpha(2m+1)}}$. Also, as $z\to\infty$, $H\to\infty$. From the curve-fitting, the value of $H_{0}$ is obtained as $\approx70$, as mentioned in Table 1. From the Hubble law $v\propto d$ i.e., $\frac{v}{d}=$constant $(H_{0})$, where $v$ is the velocity of the moving object and $d$ is the distance of moving object from the observer, increasing the value of $H_{0}$ with redshift $z$ shows the acceleration in expansion.

\begin{figure}[H]
	\includegraphics[width=9cm,height=8cm,angle=0]{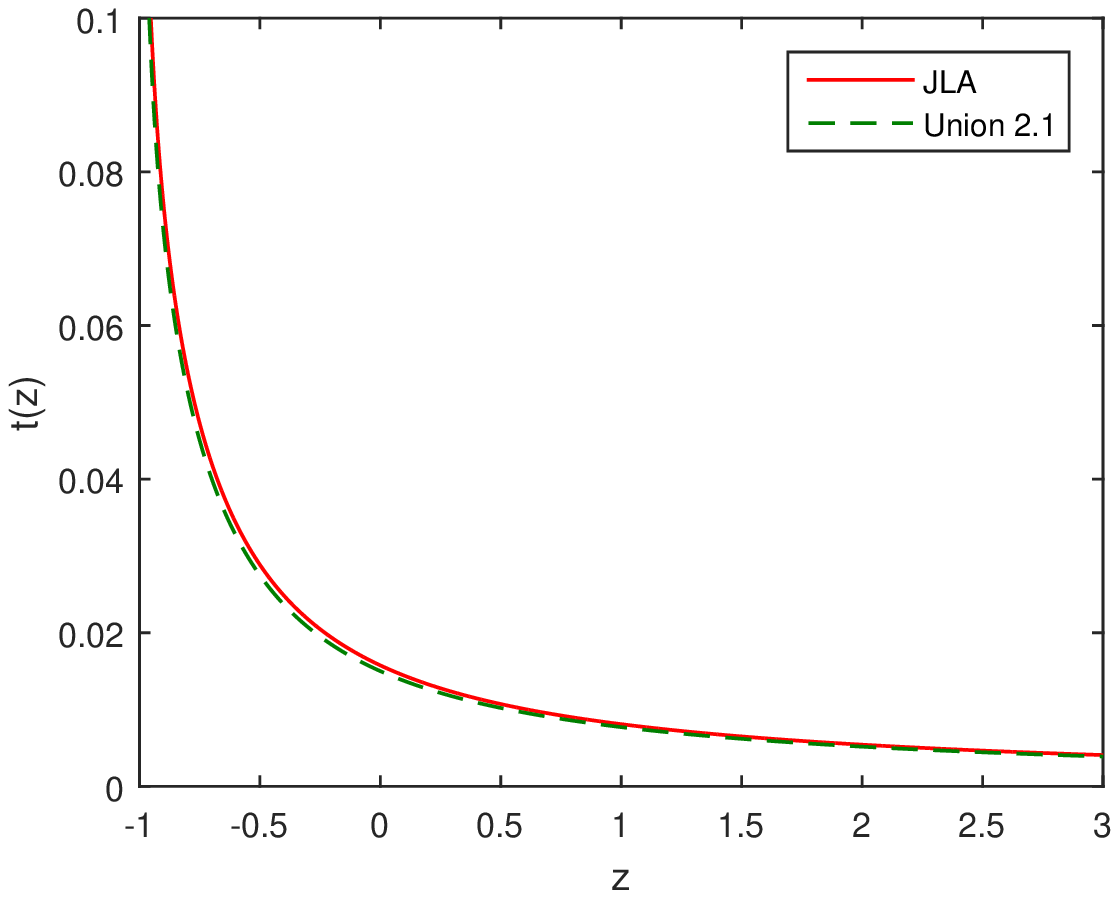}
    \includegraphics[width=9cm,height=8cm,angle=0]{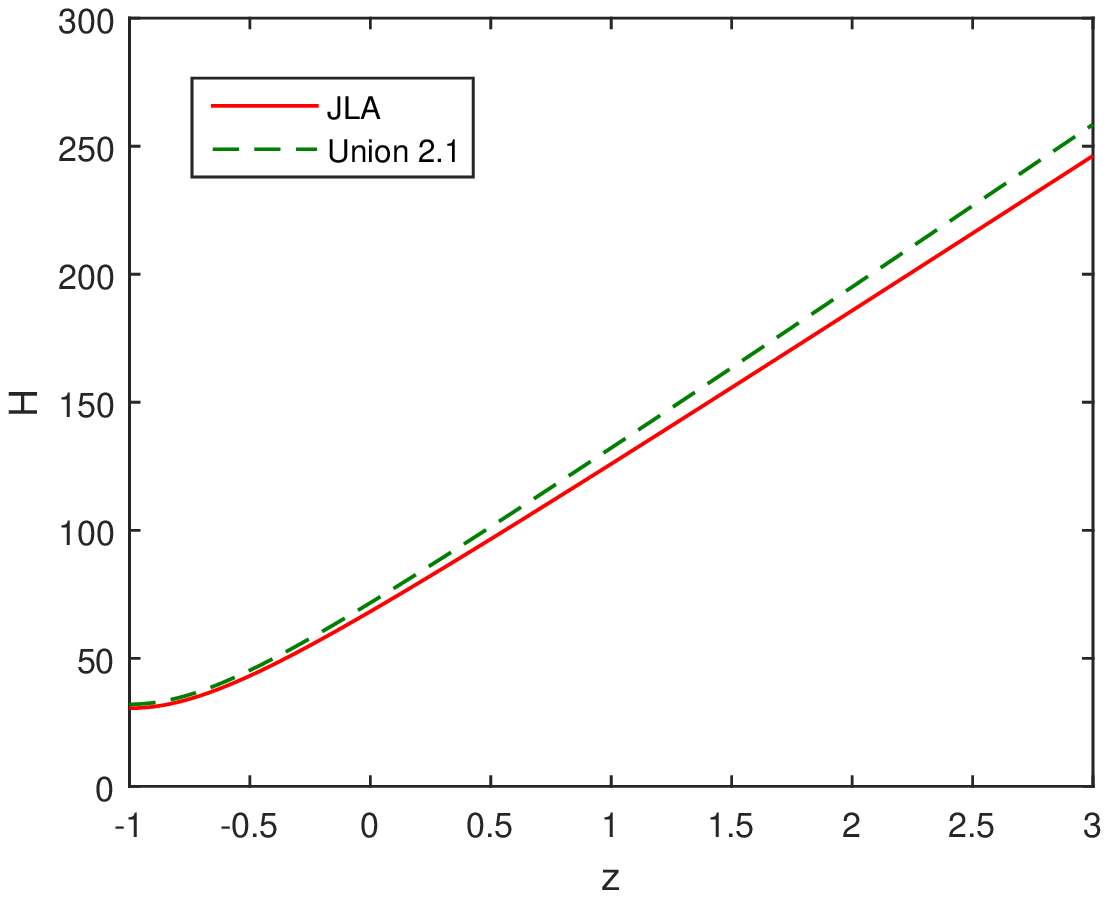}
	\caption{The graph of cosmic time $t$ and Hubble parameter $H$ for our derived model.}
\end{figure}
Equations (\ref{eq36}) \& (\ref{eq37}) represent the expressions for the deceleration parameter $q_{x}$ and $q_{y}$ along $x$ and $y$ axes respectively. Equation (\ref{eq38}) represents the expression for the average deceleration parameter $q(t)$. The geometrical behavior of these deceleration parameters is shown in Figures 5a \& 5b for two data sets, JLA and Union 2.1 compilation of SNe Ia observations. From Figures 5a \& 5b show that the values of the deceleration parameter decrease with cosmic time and as $t\to\infty$ then $q\to-1$. and it depicts that the present universe expansion is undergoing into accelerating expansion phase. We observe that the average deceleration parameter $q$ represents a universe from static to an accelerating expanding universe, but the component of $q$, along $x$-axis $q_{x}$ varies over $(-1, -0.2)$ which reveals an ever-accelerating universe and the component $q_{y}$ depicts a transit phase accelerating universe. The present value of $q_{x}$ is determined as $\{-0.4418, -0.4515\}$, $q_{y}=\{0.1164, 0.09702\}$ and $q=\{-0.1627, -0.1772\}$ along two datasets, which reveals that our present universe is undergoing an accelerating expansion phase. Equation (\ref{eq39})
represents the expression for the energy density $\rho$, and we can observe that as $t\to\infty$, then $\rho\to0$ i.e. at the time of
origin of the universe, the energy density of the universe is very high, and after Big Bang, it diluted with time and due to which universe
comes into the form of accelerating expansion. For $\rho\geq0$, we can find a constraint on $\beta$ as  $\beta\geq\frac{12\alpha(2m+1)^2}{m^4}H_{x}^{4}$.
\begin{figure}[H]
  	a.\includegraphics[width=9cm,height=8cm,angle=0]{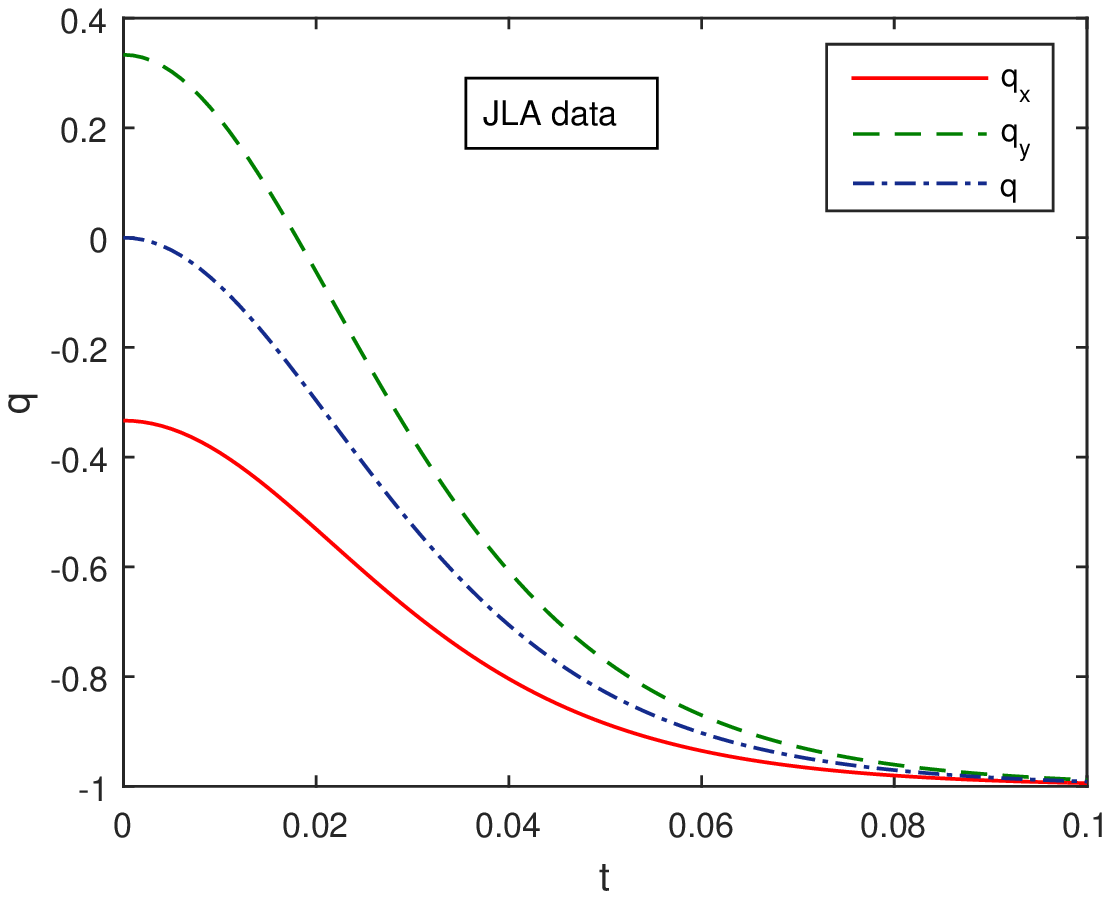}
	b.\includegraphics[width=9cm,height=8cm,angle=0]{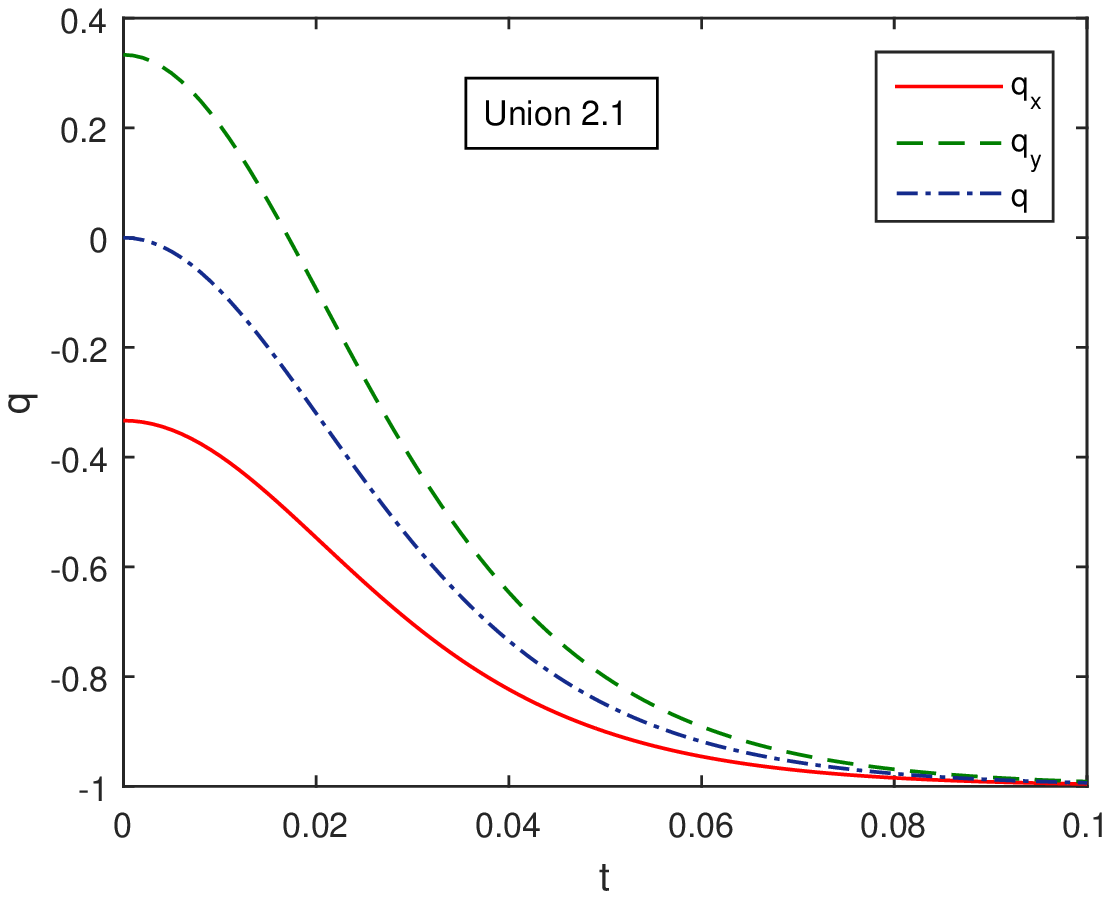}
	\caption{Behavior of deceleration parameter along two datasets.}
\end{figure}
\begin{figure}[H]
	a.\includegraphics[width=9cm,height=8cm,angle=0]{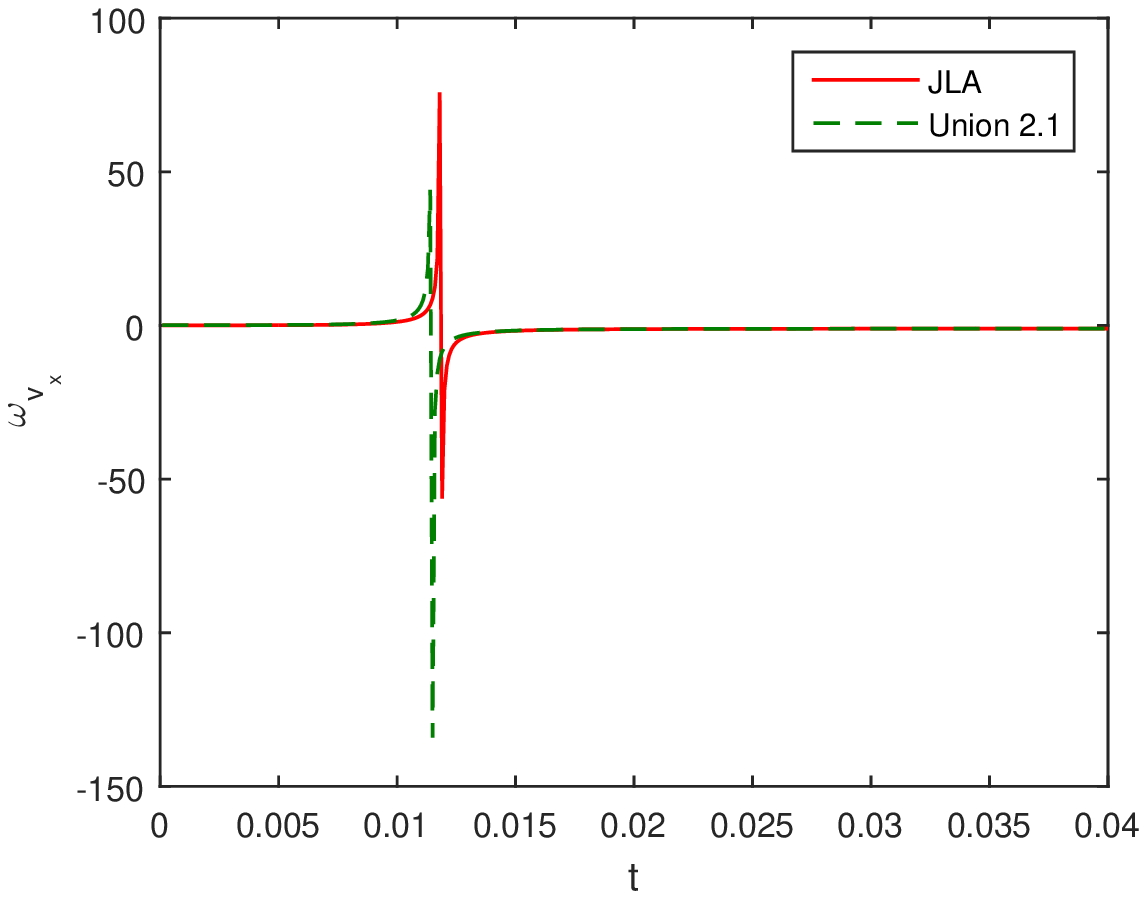}
    b.\includegraphics[width=9cm,height=8cm,angle=0]{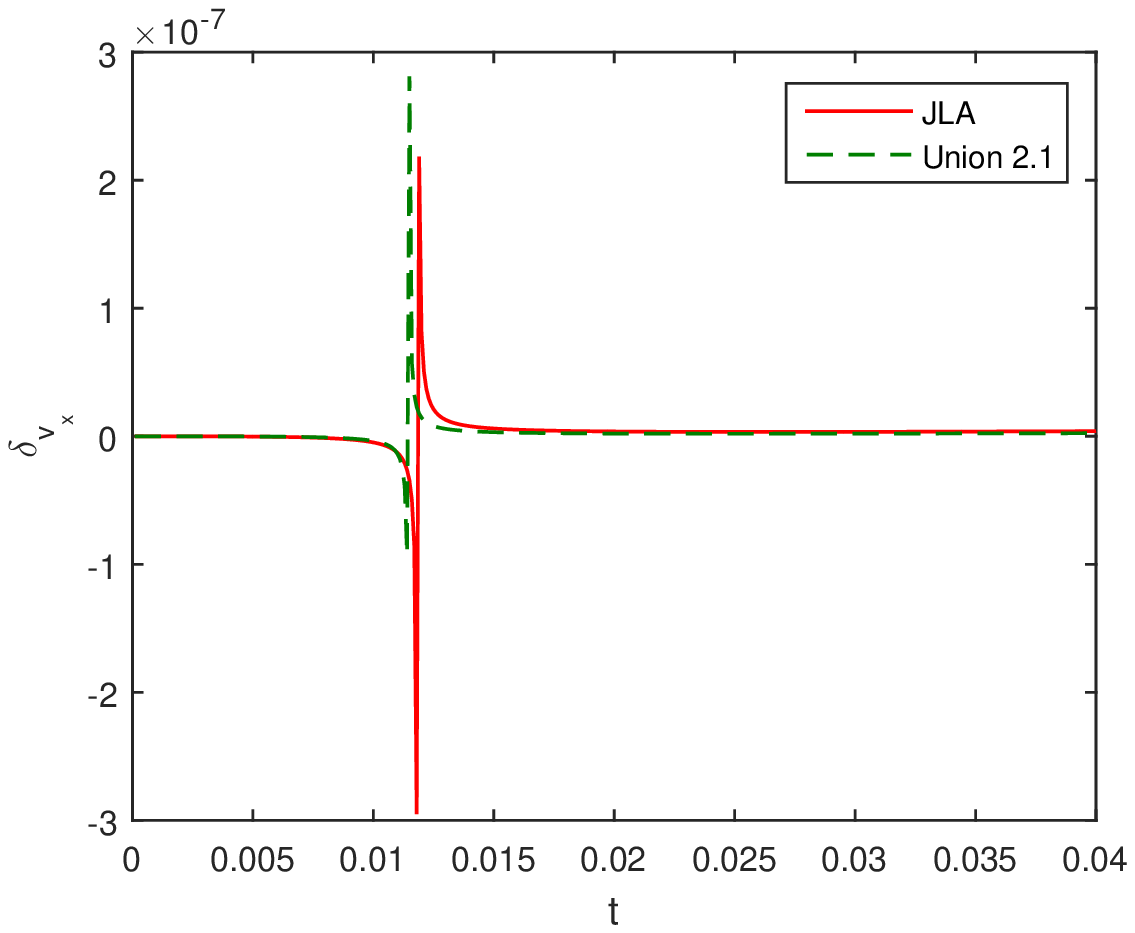}
	\caption{Variation of EoS $\omega_{v}$ and skewness $\delta_{v}$ along two datasets.}
\end{figure}

Equations (\ref{eq40}) and (\ref{eq41}) represent the expressions for equation of state parameter (EoS) $\omega_{v}$ and skewness parameter $\delta_{v}$ for bulk viscosity fluid filled in the anisotropic spacetime universe respectively and Figures 6a \& 6b show their variations over cosmic time $t$ respectively. Figure 6a shows that the EoS parameter $\omega_{v}$ is an increasing function of cosmic time $t$ with $\omega_{v}\geq0$. Then it decreases with time and tends to $-1$ ($\Lambda$CDM model) in the Late-time universe, which is a good feature of the model. It depicts
that the early universe was matter-dominated. As time passes, it is converted into dark energy, and different structures are formed in the universe. Due to this, its volume increases, expanding and accelerating the universe. Figure 6b shows that the skewness parameter $\delta_{v}$ decreases with cosmic time $t$, and the present values of $\delta_{v}$ tend to zero, which is consistent with the skewness property.\\

Equation (\ref{eq28}) with (\ref{eq35}) represents the expression for the bulk-viscosity function. We have observed that the Hubble parameter $H$ is a decreasing function of cosmic time $t$ and estimated values of $\xi_{0}=\{0.99998, 0.97455\}$ along two data sets. Hence, the bulk-viscosity function is a decreasing function of time $t$, which may cause the expanding and accelerating expanding universe.

\subsection{Age of the present universe}
The present age of the universe is estimated by
\begin{equation}\label{eq47}
t_{0}-t=-\int_{t_{0}}^{t}dt=\int_{0}^{z}\frac{dz}{(1+z)H(z)}
\end{equation}
Using (\ref{eq43}) in (\ref{eq47}) and integrating, we get
\begin{equation}\label{eq48}
t_{0}-t=\frac{6}{m+2}\sqrt{\frac{\alpha(2m+1)}{\xi_{0}}}[tanh^{-1}(\sqrt{5})-tanh^{-1}(\sqrt{1+4(1+z)^{2}})]
\end{equation}
The geometrical behaviour of $H_{0}(t_{0}-t)$ versus redshift is represented in Figure $7$. One can see that as $z\to\infty$ then $H_{0}(t_{0}-t)\to H_{0}t_{0}$ (i.e. $t\to0$), here, $H_{0}$ is the present value of $H$ and $t_{0}$ shows the age of the present universe. In our derived model, the age of the present universe is calculated as $\{13.9532, 13.2910\}$ GYrs for two data sets respectively JLA and Union 2.1 compilation of supernovae Ia. Our result is supported by
 the recent observations \cite{ref14, ref70, ref72}.
\begin{figure}[H]
	\centering
   \includegraphics[width=10cm,height=9cm,angle=0]{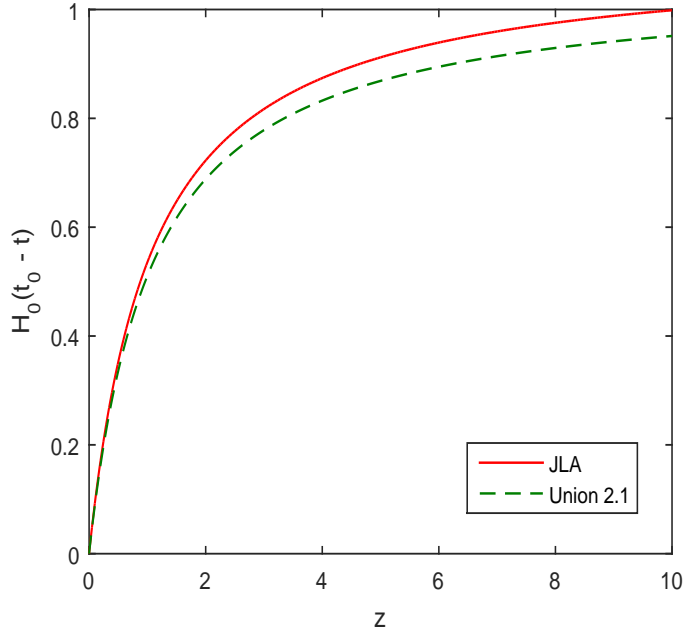}
	\caption{The graph of cosmic age of the universe in our derived model.}
\end{figure}
\subsection{Statefinder Analysis}

The Hubble parameter$H=\frac{\dot{a}}{a}$ and the deceleration parameter $q=-\frac{a\ddot{a}}{\dot{a}^{2}}$ , two cosmological parameters that characterise the geometrical evolution of the cosmos, are both known as geometrical parameters. In addition to these factors, statefinder analysis, which was proposed in reference \cite{ref14}, which depicts the various phases of evolution of the universe's dark energy models \cite{ref74,ref75,ref76}, also includes some other geometrical parameters. Two factors, $r$ and $s$, which are each expressed in terms of the average scale-factor $a(t)$, were used to get the statefinder analysis.

\begin{equation}\label{eq49}
  r=\frac{\dddot{a}}{aH^{3}},~~~~~~~~s=\frac{r-1}{3(q-\frac{1}{2})}
\end{equation}
These parameters in the current model are determined as
\begin{equation}\label{eq50}
  r=tanh^{2}\left(\frac{m+2}{6}\sqrt{\frac{\xi_{0}}{\alpha(2m+1)}}t\right)
\end{equation}
Using Eqs. (\ref{eq34}), (\ref{eq42}) and (\ref{eq43}) in Eqs. (\ref{eq49}) and (\ref{eq50}), we obtained the expression $s$ as
\begin{equation}\label{eq51}
  s=\frac{2}{3}\frac{1}{3cosh^{2}\left(\frac{m+2}{6}\sqrt{\frac{\xi_{0}}{\alpha(2m+1)}}t\right)-1}
\end{equation}
From the Eqs. (\ref{eq50}) and (\ref{eq51}), we can find that as~~ $t\to\infty$, then $r\to1$ and $s\to0$", this implies that our derived model tends to
isotropic cosmological $\Lambda$CDM model as $t\to\infty$.
\begin{figure}[H]
	\centering
	(a)\includegraphics[width=8cm,height=7cm,angle=0]{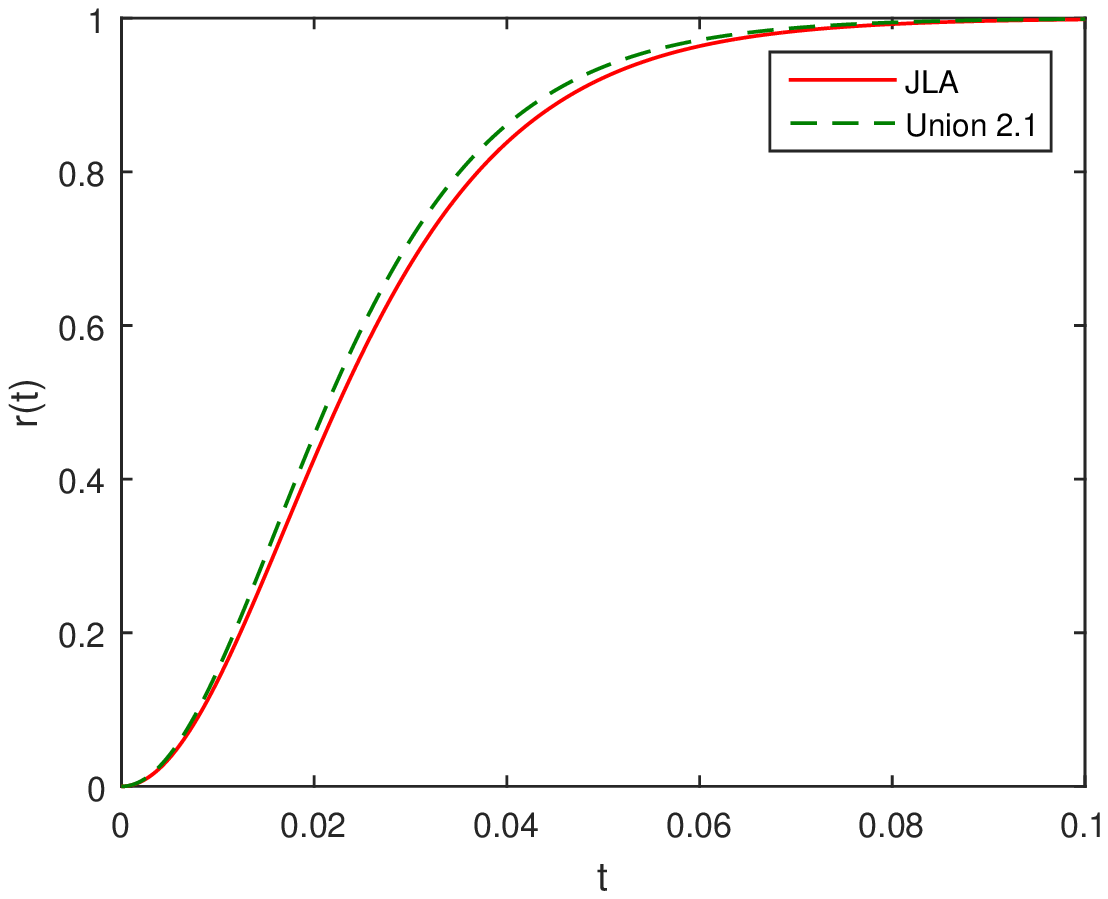}
    (b)\includegraphics[width=8cm,height=7cm,angle=0]{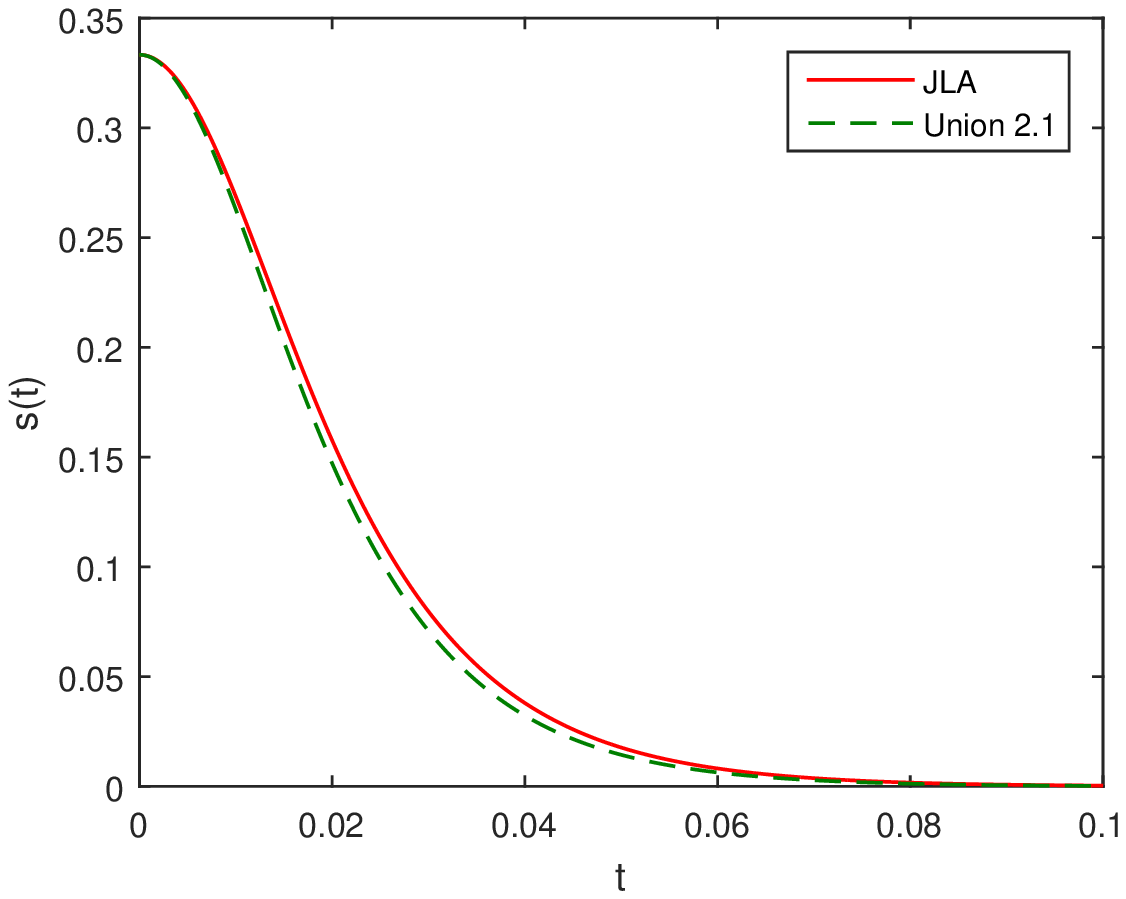}
	\caption{Behavior of statefinder parameters along two data sets.}
\end{figure}

The geometrical interpretation of the statefinder parameters $r$, $s$ with the variation of time $t$ are shown in figure 8a \& 8b respectively
and we have measured the present values of $(r-s)$ as $r_{0}=\{0.2178, 0.1999\}$, $s_{0}=\{0.2428, 0.2424\}$ respectively
for the two data sets. Also, we see that as ``$t\to\infty$ then $r\to1$ and $s\to0$" and the evolution of $(s,r)$ reveals the different stages
of dark energy models \cite{ref14, ref74, ref75, ref76}; instant, the value $(s,r)=(0,1)$ represents a flat FLRW $\Lambda$CDM model. It can
be found that at present, $(r_{0},q_{0})=(0.2178, -0.2189)$, and this indicates that the present universe is either matter-dominated or dark
energy dominated.

\subsection{Om Diagnostics}
The $O_{m}$ diagnostic is a useful tool for categorising different cosmic dark energy models \cite{ref77}. Given that it only employs the first derivative of the cosmic scale factor, this diagnosis is the most precise. It is given by for a spatially flat universe.
\begin{equation}\label{eq52}
  O_{m}(z)=\frac{-1+\left(\frac{H(z)}{H_{0}}\right)^{2}}{-1+(1+z)^{3}}
\end{equation}

where the Hubble parameter's current value is $H_{0}$. A positive slope of $Om(z)$ corresponds to phantom motion, while a negative slope denotes quintessence motion. The $\Lambda$CDM model is represented by the constant $Om(z)$.\\

In Figure 9, we found that the Om diagnostic parameter has a negative slope across a restricted set of values for the model parameter. Therefore, from the Om diagnostic test, we can conclude that the $f(Q)$ cosmological model represents quintessential behavior.\\
\begin{figure}[H]
\centering
	\includegraphics[width=10cm,height=8cm,angle=0]{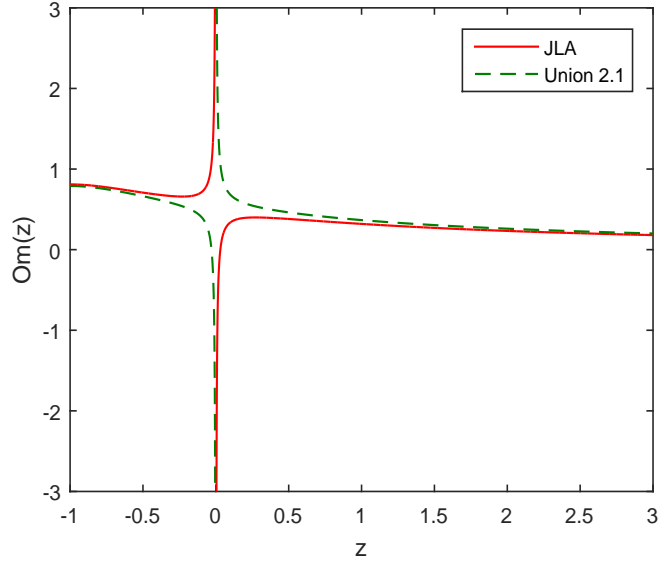}
  	\caption{Behaviour of Om diagnostics in our derived model.}
\end{figure}
\section{Concluding Remarks}
In this study, we have found a transit phase (decelerating to accelerating) cosmological model in modified $f(Q)$-gravity theory. We have investigated anisotropic scenarios of the expanding universe with viscous fluid. We solved field equations and found an exact solution as a scale factor and then analyzed the model. The main features of our model, which we observed, are as follows:

\begin{itemize}
  \item The model is with an average scale factor $a(t)$ which is obtained by solving field equations, and also, it depends upon the value of
  model parameters $\{\alpha, m, \xi_{0}\}$ is a good feature of the model.
  \item We estimated the present values of Hubble parameter $H_{0}$ as $\approx70$~Km~s$^{-1}$/Mpc and deceleration parameter $q_{0}$
  as $-0.1772$ which is supported by the recent observations \cite{ref14, ref74, ref75}.
  \item The derived model is an accelerating expanded universe and anisotropically transit phase model (decelerating to accelerating universe).

  \item We have found the present value of EoS parameter $\omega_{v}<-1/3$ for bulk-viscous fluid and it varies over
  positive to negative which indicates the existence of dark energy dominated models in the late-time universe.

  \item The skewness parameter $\delta_{v}$ for bulk-viscous fluid tends to zero which is consistent with skewness property.

  \item According to our calculations, the universe is currently $\approx13.6$ Gyrs old, which is supported by recent observations \cite{ref14, ref74, ref75}.

  \item We have found the bulk-viscosity function as a decreasing function of cosmic time $t$ which may be a dark energy candidate.

  \item The model has a singularity at $t=0$, since it is found that $a(0)=0$.
  \item The value of energy density $\rho\to\infty$ as $t\to0$ which reveals the Big-Bang singularity of the model.

  \item The analysis of the statefinder parameters indicates that the derived model tends to $\Lambda$CDM model as $t\to\infty$.

  \item In the Om diagnostic analysis of our model we have found that the derived model represents the quintessential behavior.
\end{itemize}
Thus, our derived universe model evolves with a natural scale factor and reveals the various nature of the dark energy model which will create
interests among the researchers working in this field.


\section*{Author Contributions:}
The authors contributed equally in this work. All authors have read and agreed to the published version of the manuscript.

\section*{Funding:}
This research received no external funding.

\section*{Acknowledgement}
A. Pradhan \& A. Dixit thank the IUCAA, Pune, India for providing facilities under associateship programs.

\section*{Conflicts of Interest:}
The authors declare no conflict of interest.
	
\end{document}